\documentclass[journal=cmatex,manuscript=article]{achemso}
\usepackage{chemformula} 
\usepackage[T1]{fontenc} 
\usepackage{amsmath} 
\usepackage{amssymb} 
\usepackage{graphicx} 
\graphicspath{ {./images/} }

\newcommand{\clc}{Cl@4c}
\let\oldmaketitle\maketitle
\let\maketitle\relax

\author{Jiho Lee}
\affiliation{Department of Materials Science and Engineering, Seoul National University, Seoul 08826, Korea}
\author{Suyeon Ju}
\affiliation{Department of Materials Science and Engineering, Seoul National University, Seoul 08826, Korea}
\author{Seungwoo Hwang}
\affiliation{Department of Materials Science and Engineering, Seoul National University, Seoul 08826, Korea}
\author{Jinmu You}
\affiliation{Department of Materials Science and Engineering, Seoul National University, Seoul 08826, Korea}
\author{Jisu Jung}
\affiliation{Department of Materials Science and Engineering, Seoul National University, Seoul 08826, Korea}
\author{Youngho Kang}
\affiliation{Department of Materials Science and Engineering, Incheon National University, Incheon 22012, Korea}
\author{Seungwu Han}
\email{hansw@snu.ac.kr}
\affiliation{Department of Materials Science and Engineering, Seoul National University, Seoul 08826, Korea}
\alsoaffiliation[Second University]
{Korea Institute for Advanced Study, Seoul, 02455, Korea}

\title[An \textsf{achemso} demo]
  {Disorder-dependent Li diffusion  in \ch{Li6PS5Cl} investigated by machine learning potential}


\begin{document}

\begin{tocentry}
    \includegraphics[width=3.25in]{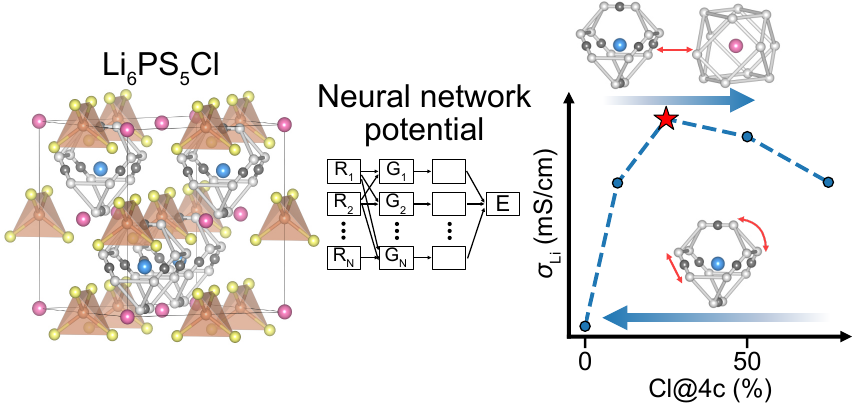}
\end{tocentry}

\oldmaketitle
\begin{abstract}
Solid-state electrolytes with argyrodite structures, such as \ch{Li6PS5Cl}, have attracted considerable attention due to their superior safety compared to liquid electrolytes and higher ionic conductivity than other solid electrolytes. Although experimental efforts have been made to enhance conductivity by controlling the degree of disorder, the underlying diffusion mechanism is not yet fully understood. Moreover, existing theoretical analyses based on {\it ab initio} MD simulations have limitations in addressing various types of disorder at room temperature. In this study, we directly investigate Li-ion diffusion in \ch{Li6PS5Cl} at 300 K using large-scale, long-term MD simulations empowered by machine learning potentials (MLPs). To ensure the convergence of conductivity values within an error range of 10\%, we employ a 25 ns simulation using a 5×5×5 supercell containing 6500 atoms. The computed Li-ion conductivity, activation energies, and equilibrium site occupancies align well with experimental observations. Notably, Li-ion conductivity peaks when Cl ions occupy 25\% of the 4c sites, rather than at 50\% where the disorder is maximized. This phenomenon is explained by the interplay between inter-cage and intra-cage jumps. By elucidating the key factors affecting Li-ion diffusion in \ch{Li6PS5Cl}, this work paves the way for optimizing ionic conductivity in the argyrodite family.

\end{abstract}

\section{1. INTRODUCTION}

With improved safety and higher energy density compared to conventional batteries, solid-state batteries employing solid Li-ion conductors are attracting attention.
\cite{10.1038/s41578-019-0165-5,
10.1038/s41563-019-0431-3,
10.1038/natrevmats.2016.103,
10.1038/s41560-023-01208-9,
10.1002/adma.201705702,
10.1021/acs.chemmater.2c01475,
10.1002/adfm.202100891}
However, the solid-state Li-ion conductors typically suffer from low ionic conductivities at room temperature. As such, a wide range of materials from oxide,
\cite{10.1021/acs.chemrev.9b00427,
10.1021/acs.chemrev.1c00594,
10.1039/C4CS00020J}
sulfide,
\cite{10.1002/adma.202000751,
10.1002/adma.201901131,
10.1002/adma.201606823}
and halide
\cite{10.1039/C9EE03828K,
10.1021/acsenergylett.2c00438}
families have been explored to increase the Li-ion conductivity to the level required in industry ($\gtrsim$ 10 mS/cm).\cite{10.1149/1945-7111/ab7f84}
Among the sulfide electrolytes, two systems have shown promising results: \ch{Li10GeP2S12} (LGPS) from the thio-LISICON family
\cite{10.1002/aenm.202002153,
10.1002/adfm.202203551}
and \ch{Li6PS5X} (X = Cl, Br) from the argyrodite family.\cite{10.1016/j.nanoen.2021.105858}
The Li-ion conductivity of LGPS is as high as 12 mS/cm at room temperature, \cite{10.1038/nmat3066}
one of the highest among the solid-state electrolytes. On the other hand, the argyrodite family has gained recent attention owing to high conductivities ($\sim$5 mS/cm),
\cite{10.1021/acs.chemmater.0c02418,
10.1021/jacs.7b06327,
10.1021/acsami.8b07476,
10.1021/acs.chemmater.0c04650}
and lower material cost than LGPS. \cite{10.1002/batt.202200553}
In addition, through doping with other halogen or group IV and V elements such as Si, Ge, Sn, and Sb, ionic conductivities exceeding 10 mS/cm and wide electrochemical windows were achieved.
\cite{10.1002/batt.202200553,
10.1021/acs.nanolett.9b04597,
10.1002/anie.201814222,
10.1021/jacs.9b08357,
10.1021/acssuschemeng.0c05549}

\begin{figure*}
  \includegraphics[width=6in]{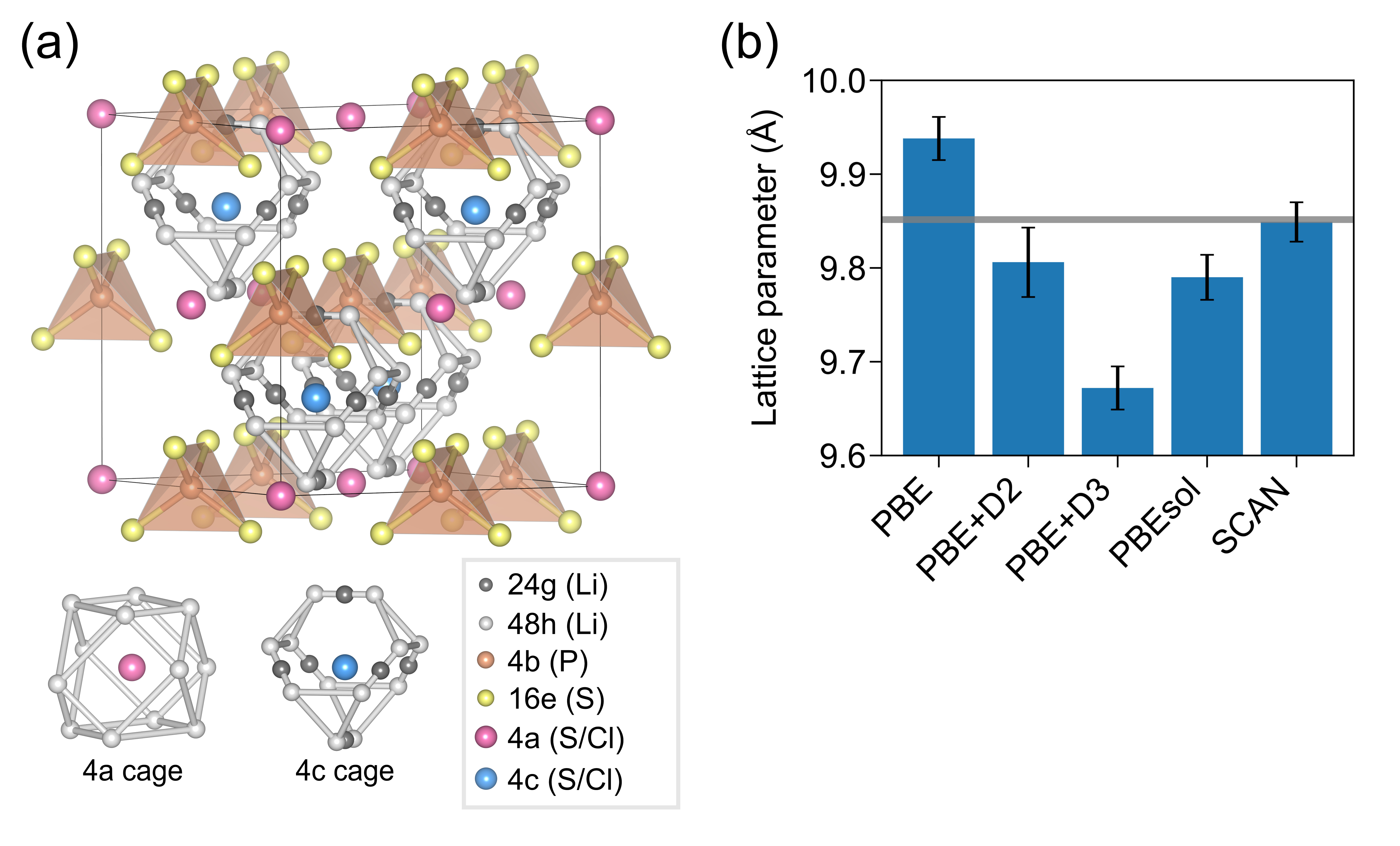}
  \caption{(a) Structure of argyrodite \ch{Li6PS5Cl}.  The S and Cl atoms could exist in both 4a and 4c sites, resulting in site disorders. (b) Lattice parameters computed with different functionals on 50\% Cl@4c unit cells. The horizontal gray band indicates the experimental range of lattice parameters and the error bar shows the standard deviation of the lattice parameter values.}
  \label{fgr:structure_functional}
\end{figure*}

The high conductivity in the argyrodite family has been suggested to be due to anion disorder. For instance, \ch{Li6PS5I} has significantly lower conductivities ($\sim$1 $\mu$S/cm) than \ch{Li6PS5Cl} or \ch{Li6PS5Br} (1–10 mS/cm),\cite{10.1021/jacs.7b06327, 10.1002/anie.200703900,10.1039/c9ta02126d,10.1016/j.jallcom.2018.03.027} although they all share the same crystal structure (see Figure~\ref{fgr:structure_functional}). Diffraction measurements revealed that while I atoms fully occupy 4a sites without any disorder, in \ch{Li6PS5Cl} or \ch{Li6PS5Br}, Cl atoms randomly occupy 4c sites by 50\%–60\%, Br atoms do so at a rate of 15\%–25\%, and S atoms, in turn, occupy 4a sites by the corresponding amounts.
\cite{10.1021/jacs.7b06327,
10.1021/acs.inorgchem.0c01504,
10.1021/acs.inorgchem.8b02443}
The substantial difference in the ionic radii of S and I hinders the mixing of anions while similar radii of S, Cl, and Br lead to site disorders. (The ionic radii are 1.84 Å for \ch{S^{2-}}, 1.81 Å for \ch{Cl-}, 1.96 Å for \ch{Br-}, and 2.20 Å for \ch{I-} ions, respectively.\cite{10.1107/s0567739476001551})

The \textit{ab initio} molecular dynamics (MD) simulation based on the density functional theory (DFT) has proven useful for investigating Li diffusion in the argyrodite subjected to various chemical modifications. For instance, the diffusion mechanism and the role of anion disorder at 4a/4c sites were explored in refs~\citenum{10.1039/d1ta10964b, 10.1021/acs.chemmater.0c03738, 10.1039/d2ra05889h, 10.1021/acs.chemmater.9b02047, 10.1002/aenm.202003369}, revealing that the size uniformity of the 4a and 4c cages closely relates to the Li-ion conductivity. In addition, new compositions \cite{10.1021/acs.chemmater.0c04650,
10.1021/acs.chemmater.6b03630,
10.1021/acs.chemmater.6b02648,
10.1016/j.ensm.2020.04.042}
and doped or alloyed systems \cite{10.1021/acssuschemeng.0c05549, 10.1002/eem2.12282}
were investigated by DFT, which supported experimental measurements and suggested promising engineering directions. However, from a quantitative prediction standpoint, current computational simulations have been found to be unsatisfactory. For instance, the activation energies for Li diffusion in \ch{Li6PS5Cl} were computed to be 0.18–0.24 eV
\cite{10.1039/d2ra05889h,
10.1021/acs.chemmater.9b02047,
10.1016/j.ensm.2020.04.042}, significantly lower than the experimental data of 0.27–0.34 eV
under similar degrees of disorder\cite{10.1021/acsami.8b07476,
10.1021/acsami.8b15121,
10.1149/2.0301903jes}.
More significantly, the estimated conductivities at 300 K vary widely among literature (6–115 mS/cm)
\cite{10.1038/s41524-020-00432-1,10.1039/d1ta10964b,
10.1039/d2ra05889h,
10.1021/acs.chemmater.6b03630}
and are considerably higher than the experimental measurements (2–5 mS/cm).
\cite{10.1021/acs.chemmater.0c02418,
10.1021/jacs.7b06327,
10.1021/acsami.8b07476,
10.1021/acs.chemmater.0c04650}
Given that DFT studies on LGPS \cite{10.1021/cm203303y,10.1039/c2ee23355j}
and \ch{Li7La3Zr2O12} \cite{10.1038/s41524-018-0074-y,10.1021/cm303542x}
have produced ionic conductivities in good agreements with experimental results, the above discrepancy for \ch{Li6PS5Cl} warrants further investigation. Moreover, for the prediction and understanding of material engineering such as doping, there is a pressing need for a reliable computational framework.

Two factors may have contributed to the significant discrepancy between DFT and experiment highlighted above: i) Most of the previous studies employed the PBE functional, which tends to overestimate lattice parameters, consequently leading to an overestimation of conductivities. ii) Moreover, many studies utilized unit cells to evaluate conductivity.
\cite{10.1039/d1ta10964b,
10.1039/d2ra05889h,
10.1021/acs.chemmater.9b02047,
10.1021/acs.chemmater.6b03630,
10.1021/acs.chemmater.6b02648,
10.1002/eem2.12282}
While a lattice parameter of 1 nm of the unit cell would be sufficient to prevent direct chemical interactions between periodic images, such small supercells are susceptible to spurious correlated Li-ion diffusion between neighboring cells and are also constrained in adequately considering S/Cl disorder. \cite{10.1039/d1ta10964b}
The first issue can be resolved by adopting a more sophisticated functional, but the second problem necessitates the use of large-scale, long-term simulations, which would be prohibitively expensive for DFT calculations. 

Recently, machine learning potentials (MLPs) are attracting attention as they can address the limitations of DFT in time- and size-scales. As such, MLPs have been actively applied to solid-state electrolytes.
\cite{10.1088/2752-5724/acb506,
10.1016/j.mtphys.2021.100463,
10.1063/5.0090341,
10.1063/5.0041849,
10.1088/2515-7655/acbbef}
For instance, MLP-MD simulations were able to identify transitions between quasi-linear Arrhenius regimes at room temperature in materials such as \ch{Li_{0.33}La_{0.56}TiO3}, \ch{Li3YCl6}, \ch{Li7P3S11}, and LGPS. \cite{10.1016/j.mtphys.2021.100463,10.1088/2515-7655/acbbef}
The transitions arise from the activation of additional diffusion pathways at high temperatures, which is challenging to observe with DFT due to the rare occurrence of Li-ion jumps at low temperatures. 
In \ch{Li7La3Zr2O12} and LGPS-type superionic conductors,  MLPs were employed to minimize statistical fluctuations. \cite{10.1063/5.0090341,10.1063/5.0041849}
For the argyrodite systems, moment tensor potential had been generated for \ch{Li6PS5Cl} and \ch{Li6PS5I}. \cite{10.1088/2752-5724/acb506}
However, the simulations were performed at high temperatures without considering disorder effects. 

Addressing the above issues, we herein evaluate the ionic conductivity of the archetypal argyrodite system \ch{Li6PS5Cl} using Behler-Parrinello-type neural network potentials (NNPs)\cite{10.1103/physrevlett.98.146401}.
Harnessing the speed of NNP, we evaluate theoretical Li conductivity of \ch{Li6PS5Cl} directly at room temperature. 
To obtain the conductivity which is statistically converged with an error range of 10\%, we carry out systematic tests on the supercell and ensemble sizes. The computed ionic conductivity and activation energy of Li-ion diffusion are comparable to the experimental data. The diffusion mechanism at 300 K is analyzed in detail to explain the disorder-dependent Li conductivity and identify key factors.

\section{2. METHODS}

\subsection{2.1. Crystal Structure and DFT Parameters}

Figure~\ref{fgr:structure_functional} shows the unit cell of argyrodite \ch{Li6PS5Cl}. In detail, P at 4b sites and S at 16e sites form a \ch{PS4} tetrahedral backbone where every site is fully occupied with one species. On the other hand, 4a and 4c sites are randomly shared by S and Cl. In this work, we specify the random occupancy using the notation of Cl@4c. For instance, 25\% Cl@4c means that Cl atoms occupy 25\% and S atoms occupy 75\% of the 4c sites, respectively, while the 4a sites are occupied by the reversed percentages. Li atoms occupy 48h sites with the site occupancy of 0.5, forming Li cages around 4a or 4c sites (see bottom figures in Figure~\ref{fgr:structure_functional}a). Lastly, 24g sites are instantly accessed during rapid shuttling motions of Li within the 48h-24g-48h doublet.

As mentioned in the introduction, the lattice parameter critically affects the calculated ionic conductivities so we first compare various functionals in terms of the equilibrium lattice parameter to determine the functional in generating training sets. The DFT calculations throughout this work are carried out using the Vienna $ab$ $initio$ simulation package (\texttt{VASP}).\cite{10.1103/physrevb.59.1758} The cutoff energy for the plane-wave basis set is chosen to be 300 eV with the 2$\times$2$\times$2 {\bf k}-point grid for the unit cell, which ensures convergence of the energy, atomic forces, and stress components to within 5 meV/atom, 0.03 eV/\AA, and 3 kbar, respectively. As the exchange-correlation functionals, we consider the generalized gradient approximation (GGA)\cite{10.1103/physrevlett.77.3865} by Perdew-Burke-Ernzerhof (PBE), adding D2\cite{10.1002/jcc.20495} or D3\cite{10.1063/1.3382344, 10.1002/jcc.21759} for the van der Waals functional, PBEsol,\cite{10.1103/physrevlett.100.136406} and strongly constrained and appropriately normed (SCAN) functional.\cite{10.1103/physrevlett.115.036402} For the partial occupancy of S/Cl configurations at 4a/4c sites, we select 50\% Cl@4c, which lies in the experimental range (50\%–60\%).\cite{10.1021/acs.chemmater.0c02418, 10.1021/acs.inorgchem.0c01504, 10.1021/acs.chemmater.9b01550}
Within the unit cell structure, there are two possible S/Cl configurations that correspond to 50\% Cl@4c. For each configuration, 24 Li distributions are generated  by randomly placing one Li atom at either 48h sites in the 48h-24g-48h unit. The lattice parameter and atomic positions are then relaxed while maintaining the cubic cell shape until the remaining atomic forces and hydrostatic pressures become smaller than 0.02 eV/\AA\ and 10 kbar, respectively.

Figure~\ref{fgr:structure_functional}b compares  lattice parameters of the tested functionals, which are averaged over 50 structures (25 random Li distributions multiplied by two S/Cl configurations). The actual numbers are 9.938 \AA\ (PBE), 9.806 \AA\ (PBE-D2), 9.672 \AA\ (PBE-D3), 9.790 \AA\ (PBEsol), and 9.849 \AA\ (SCAN). The standard deviations are marked by error bars, which range over 0.023–0.037 \AA. The experimental value of 9.848–9.855 \AA\ with 53.8\%–61.5\% Cl@4c are displayed by a horizontal gray band.\cite{10.1021/acs.chemmater.0c02418, 10.1021/acs.inorgchem.0c01504, 10.1021/acs.chemmater.9b01550} It is seen that the lattice parameter of 9.849 \AA\ obtained by the SCAN functional best aligns with the experiment. Therefore, we employ the SCAN functional in constructing the training set. 

\subsection{2.2. Training Set}

The training set is constructed from 2×2×2 supercells of \ch{Li6PS5Cl} with 0\%, 50\%, and 100\% Cl@4c. For 50\% Cl@4c, two structures with random S/Cl distributions are considered. The Li atoms are initially placed at the 24g sites, and the structures are relaxed until the remaining atomic forces and hydrostatic pressures become smaller than 0.02 eV/\AA\ and 1 kbar, respectively. (During relaxation, Li atoms shift to 48h sites.) To account for volumetric changes and increase the diversity in local structures, the structures deformed by small shear strains are also considered. \textit{Ab initio} molecular dynamics (AIMD) simulations are conducted on the two 50\% Cl@4c supercells, using NVT ensembles at temperatures of 600 and 1200 K, which is controlled by a Nosé-Hoover thermostat. The total simulation time is 10 ps with a timestep of 2 fs, and training structures are sampled every 20 fs after the initial 300 fs. In total, we construct a training set with 1,995 structures and 829,920 local atomic environments (Table S1).

One might question whether the S/Cl disorder is adequately represented by the current training set, which consists of 2×2×2 supercells with 50\% Cl@4c. Within the cubic symmetry of \ch{Li6PS5Cl}, there are four equivalent 4a sites adjacent to each 4c site, and vice versa (see Figure 1a). We find that the training set encompasses 85\% of all possible S/Cl occupations at the neighboring 4a/4c sites (out of 20 combinations in total), and the omitted cases are statistically less frequent (for example, S atoms surrounded solely by other S atoms). Given that inter-cage Li hopping occurs between adjacent 4a and 4c cages, the current training set has effectively sampled the chemical environments that are crucial for Li diffusion.

\subsection{2.3. NNP Training}

We generate a Behler-Parinello-type NNP\cite{10.1103/physrevlett.98.146401} using the \texttt{SIMPLE-NN} program.\cite{10.1016/j.cpc.2019.04.014} For each element, 8 radial and 18 angular components for each element are used for symmetry function vectors, with the cutoff distance of 6 \AA. (The hyperparameters for atom-centered symmetry functions are listed in Table S2.) The 30-30 hidden layers are used in multi-layer perceptrons. We use a learning rate of 0.0001, and the training data are divided into a 9:1 ratio for the training and validation sets. The learning speed is notably enhanced by applying principal component analysis to the input vectors and subsequently whitening them.\cite{10.1016/j.commatsci.2020.109725} The loss function ($\Gamma$) consists of energy, force, and stress terms with an L2 regularization coefficient of $10^{-6}$ to avoid overfitting of the model:
\begin{align}
\begin{split}
\Gamma &= \frac{1}{M} \sum_{i=1}^{M} \left( \frac{E_i^{\mathrm{DFT}}-E_i^{\mathrm{NNP}}}{N_i} \right) ^2 \\
&+      \frac{\mu_1}{3\sum_{i=1}^{M} N_i} \sum_{i=1}^{M} \sum_{j=1}^{N_i} \left| \mathbf{F}_{ij}^{\mathrm{DFT}} - \mathbf{F}_{ij}^{\mathrm{NNP}} \right|^2 \\ 
&+      \frac{\mu_2}{6M} \sum_{i=1}^{M} \sum_{k=1}^{6} \left| S_{ik}^{\mathrm{DFT}} - S_{ik}^{\mathrm{NNP}} \right|^2
\end{split}
\end{align}
where $M$ the total number of structures in the training set, $N_i$ the number of atoms in the $i$th structure, $E_i^{\mathrm{DFT(NNP)}}$, $\mathbf{F}_{ij}^{\mathrm{DFT(NNP)}}$, and $S_{ik}^{\mathrm{DFT(NNP)}}$ are the total energy, atomic force of the $j$th atom, the $k$th component of the (virial) stress tensor obtained by DFT (NNP) calculations, respectively. The parameters $\mu_1$ $( = 0.1)$ and $\mu_2$ $(=10^{-6})$ are used to scale the relative importance of atomic force and stress to the total energy in minimizing the loss function. 

We train an NNP with energy, force, and stress root mean square errors (RMSEs) of 1.16 meV/atom, 0.13 eV/\AA, and 1.08 kbar for the validation set, respectively. (See Figure S1 for the parity plot of energy and forces.) Throughout this work, NNP-MD simulations are carried out with the \texttt{LAMMPS} package~\cite{10.1016/j.cpc.2021.108171} with a timestep of 2 fs and the Nosé-Hoover thermostat for NVT ensemble.

\subsection{2.4. Ionic Conductivity and Activation Energy}

The mean squared displacement (MSD) is obtained by averaging the displacements of mobile atoms generated by MD simulations over time duration $t$:
\begin{equation}
\mathrm{MSD}(t) = \frac{1}{N_{\mathrm{Li}}} \sum_{i}^{N_{\mathrm{Li}}} |\mathbf{r}_i(t)-\mathbf{r}_i(0)|^2
\end{equation}
where $N_{\mathrm{Li}}$ is the number of Li ions in the supercell and $\mathbf{r}_i(t)$ is the position of the $i$th Li ion at time $t$.
The self diffusivity ($D_{\mathrm{Li}}$) is obtained from the slope of MSD-time curve considering the dimensionality ($d = 3$). \cite{10.1063/1.1777148}
\begin{equation}
D_{\mathrm{Li}} = \lim_{t\to\infty}\frac{\mathrm{MSD}(t)}{6t}
\end{equation}
The Li conductivity ($\sigma_{\mathrm{Li}}$) is then calculated from a given temperature based on the Nernst-Einstein equation: \cite{10.1063/1.1777148}
\begin{equation}
\label{eqn:Nernst-Einstein}
\sigma_{\mathrm{Li}} = \frac{N_{\mathrm{Li}} z^2 e^2}{V k_{\rm B} T}D_{\mathrm{Li}}
\end{equation}
where $V$ the supercell volume, $e$ the electron charge, $z = 1$ for the charge of Li ions, $k_{\rm B}$ the Boltzmann constant, and $T$ is the temperature.

The Arrhenius-type representation of diffusivity is as follows:
\begin{equation}
\label{eqn:Arrhenius}
D_{\mathrm{Li}} = D_0 \exp \left( -\frac{E_{\mathrm a}}{k_{\mathrm B} T} \right)
\end{equation}
where $D_0$ is the pre-exponential factor and $E_{\mathrm a}$ is the activation energy of diffusion. Inserting eq~\ref{eqn:Arrhenius} into eq~\ref{eqn:Nernst-Einstein} gives:
\begin{equation}
\log (\sigma_{\mathrm{Li}} T) = - \frac{E_{\mathrm a} }{k_{\mathrm B} T} + C
\end{equation}
where $C$ is a temperature-independent constant. Thus, one can obtain $E_{\mathrm a}$ from the slope of $1/T$ versus $\log(\sigma_{\mathrm{Li}} T)$ plot.

\section{3. RESULTS AND DISCUSSION}

\subsection{3.1. Validation of Neural Network Potential}

\begin{figure*}
  \includegraphics[width=6in]{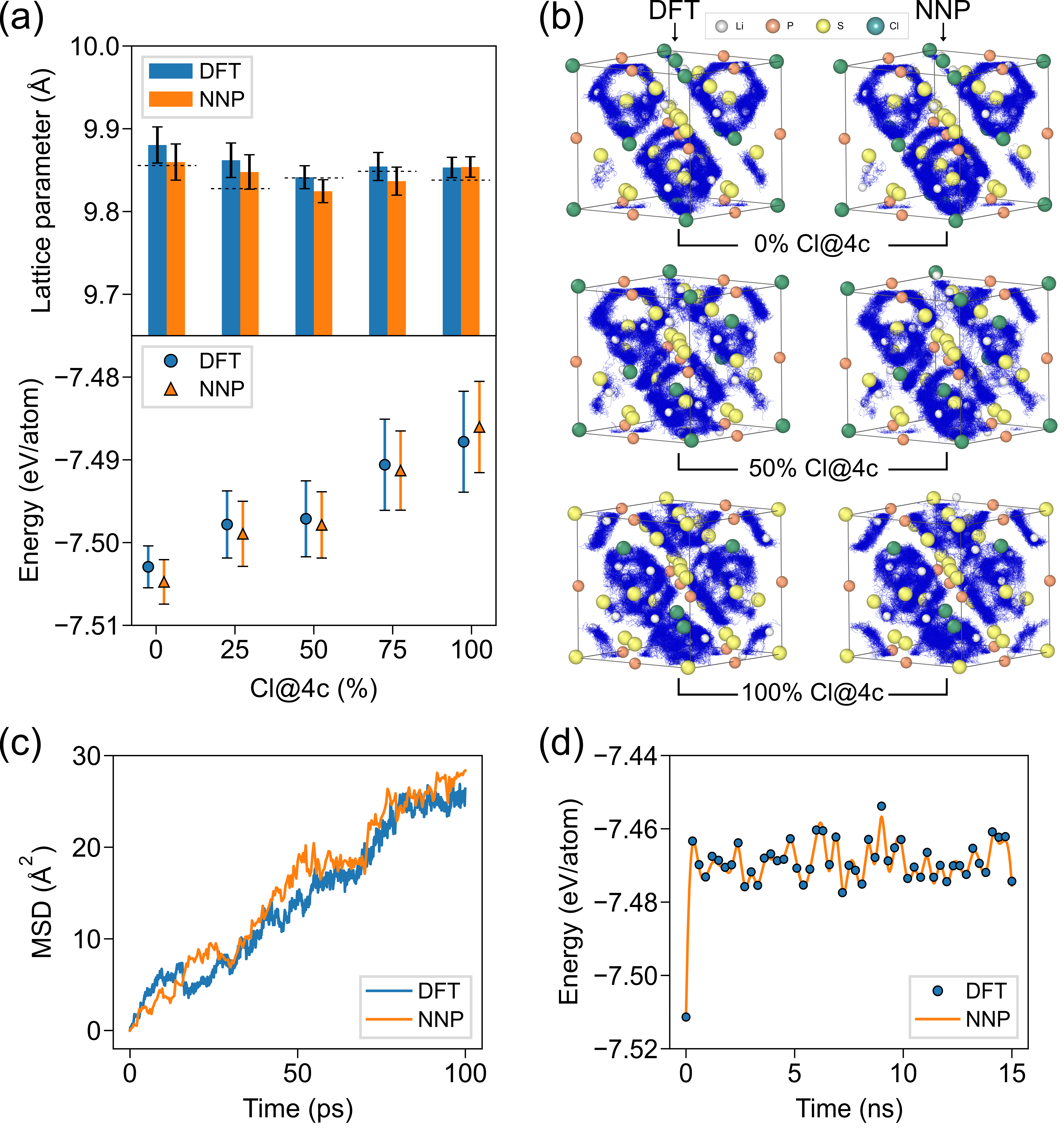}
  \caption{Validation of NNP for the unit cells. (a) The upper half depicts the lattice parameters of 0\%, 25\%, 50\%, 75\%, and 100\% Cl@4c calculated with DFT and NNP. The lower half shows the energies of unit cells per atom. Blue and orange colors represent DFT and NNP values, respectively. The error bars indicate the standard deviation of 25 structures each. The black dashed lines indicate the lattice parameters of the structures with the lowest DFT energy in each Cl@4c. (b) Trajectories of Li atoms in unit cells with 0\%, 50\%, and 100\% Cl@4c, calculated with DFT (left column) and NNP (right column) at 500 K. (c) MSD plot of both DFT and NNP at 500 K MD simulations in 50\% Cl@4c. (d) Energy plot of NNP-MD simulation and DFT single-point calculations on the MD trajectory snapshots in 50\% Cl@4c.}
  \label{fgr:validation}
\end{figure*}

Before performing large-scale simulations with NNP, we further validate the trained potential by comparing it with DFT results for systems that were not explicitly included in the training set. To this end, we prepare unit-cell models with specific S/Cl configurations for 0\%, 25\%, 50\%, 75\%, and 100\% Cl@4c. First, to examine static properties, we consider 25 structures for each Cl@4c with various Li distributions at the 48h sites. The lattice parameters and atomic positions are independently relaxed using both DFT and NNP, following the same approach outlined in Section 2.1. Figure~\ref{fgr:validation}a compares the average lattice parameters and  energies per atom between DFT and NNP after relaxation, together with standard deviations among 25 structures displayed in error bars.
The magnitude of errors relative to DFT is less than 0.03\% for the lattice parameter and 0.3\% for energy, respectively, which confirms sufficient accuracy of the trained NNP. It should be noted that 25\% and 75\% Cl@4c configurations were not included in the training set, but the corresponding results still maintain the same level of accuracy.

Next, for dynamical tests, we carry out 100-ps MD at 500 K with various \clc. Here we use the same lattice parameters for both DFT and NNP that correspond to the lowest DFT energy among 25 structures, which are indicated by dashed  lines in Figure 2a. 
Figure~\ref{fgr:validation}b displays Li trajectories during MD for 0\%, 50\%, and 100\% Cl@4c. The trajectories are visualized by placing blue dots at Li positions every 10 fs, starting from the same Li distributions for both DFT and NNP. It is seen that the Li trajectories are similar between DFT and NNP, commonly sharing characteristic features. To be specific, at 0\% and 100\% Cl@4c, Li atoms are mostly confined to 4c and 4a cages (see Figure~\ref{fgr:structure_functional}a), respectively, and inter-cage hopping is rarely observed. In these limiting cases, S$^{2-}$ ions fully occupy 4c or 4a sites and attract Li ions more strongly than Cl$^-$ ions at neighboring sites, and so inter-cage hopping between 4a and 4c cages is suppressed, resulting in effectively isolated Li cages. It is also visually noticeable that 4c cages are more compact than 4a cages, consistent with ref~\citenum{10.1039/d1ta10964b}.
On the other hand, at 50\% Cl@4c, S atoms partially occupy both 4a and 4c sites, facilitating inter-cage Li jumps.
In Figure~\ref{fgr:validation}c, we plot  MSD($t$) of Li ions from trajectories  for 50\% \clc . (The MSD curves for other values of \clc\ are shown in Figure S2.) The similar behaviors are observed again, consistent with Figure~\ref{fgr:validation}b.

The dynamical tests in Figures 2b and 2c were carried out at 500 K. Since the potential energy surface (PES) relevant for Li diffusion at  room temperature could be different from those at higher temperatures, it is worthwhile to validate the NNP directly at 300 K. To this end, we carry out a 15-ns MD at 300 K using NNP and compare its energy with DFT results for the snapshots sampled every 300 ps. Figure 2d shows the results for 50\% Cl@4c. (Other cases are displayed in Figure S3.) It can be seen that the DFT and NNP energies agree well, with the RMSE being 1.02 meV/atom, close to the RMSE for the validation set (1.16 meV/atom). The extensive comparison with DFT in this subsection confirms that the NNP produces PES close to that by DFT.

\subsection{3.2. Computational Parameters for Statistically Converged Li Conductivity}

Ideally, if one can simulate a macroscopically large-scale model over an extended period, such a simulation could address the randomness of S/Cl disorders as well as ergodicity in Li diffusion. This would yield a unique, statistically converged  $\sigma_{\mathrm{Li}}$ at all temperatures. Although NNP can handle much larger systems and longer MD simulation times compared to DFT, it still falls short when attempting to simulate systems on macroscopic scales. Therefore, it is necessary to conduct systematic convergence tests to identify a minimal simulation setup that is computationally manageable. 
Specifically, we examine how MSD and $\sigma_{\rm Li}$ depend on the supercell size, total simulation time, and initial conditions such as Li distributions. 
To this end, we perform 25-ns MD simulations at 300 K, incrementing the supercell multiplicity from 1×1×1 up to 5×5×5. For each supercell, S/Cl atoms are randomly placed within 50\% Cl@4c, and five simulations are performed independently by varying the initial velocities and Li distributions at 48h sites.
The lattice parameter in each simulation is determined independently in the following manner: First, a 1-ns MD simulation is carried out to establish a quasi-equilibrium, and full structural relaxations are carried out while retaining the cubic shape. The mean lattice parameters are 9.798, 9.803, 9.805, 9.805, and 9.804 \AA\ for 1×1×1 up to 5×5×5 supercells, respectively, with standard deviations  for each size smaller than 0.002 \AA. 

Figure 3a displays the MSD for each supercell, averaged over the five runs, with standard deviations indicated by semi-transparent shades. 
It is evident that the MSD for the 1×1×1 supercell, which has been used in many previous studies,\cite{10.1039/d1ta10964b,
10.1039/d2ra05889h,
10.1021/acs.chemmater.9b02047,
10.1021/acs.chemmater.6b03630,
10.1021/acs.chemmater.6b02648,
10.1002/eem2.12282} is significantly larger than that for larger supercells. Even though the unit cell length is close to 10 \AA\ -- long enough to avoid direct chemical interactions -- the Li jumps are correlated due to periodic boundary conditions, thereby significantly increasing the MSD. An overestimation of $\sigma_{\mathrm{Li}}$ in smaller supercells has also been reported for \ch{Li10GeP2S12}-type superionic conductors.~\cite{10.1063/5.0041849} As a result, we exclude the 1×1×1 supercell from further discussions  below.

\begin{figure*}
  \includegraphics[width=6in]{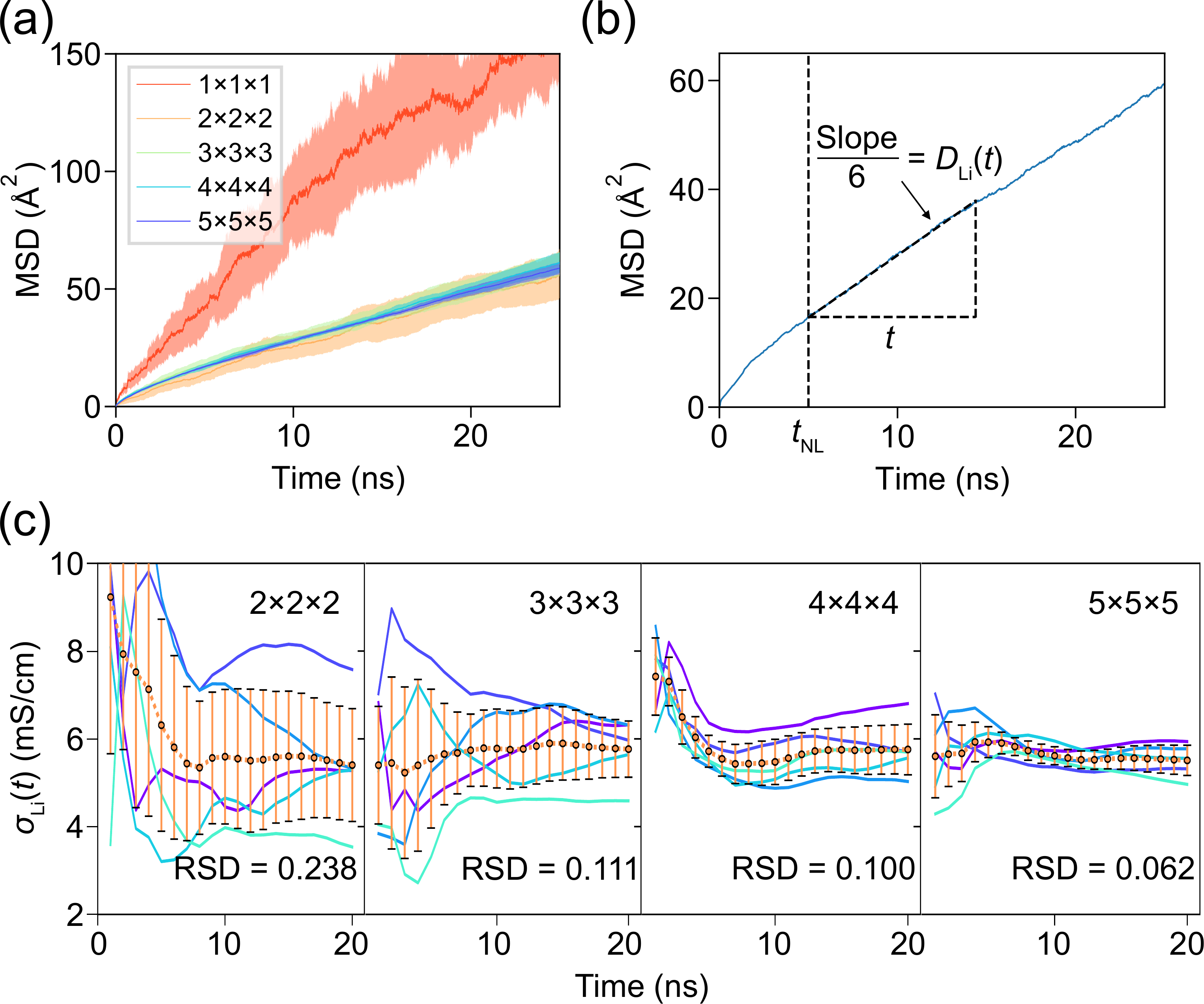}
  \caption{Parameter tests for obtaining statistically converged $\sigma_{\mathrm{Li}}$. (a) The MSD($t$) of 1×1×1 up to 5×5×5 supercells at 50\% Cl@4c are depicted with red, orange, green, blue and purple, respectively. The lines indicate the average MSD of 5 simulations, and shades represent the standard deviations. (b) The blue line shows MSD($t$) of a 50\% Cl@4c 5×5×5 supercell simulation performed at 300 K. The vertical dashed line divides the nonlinear and linear regions of the MSD($t$). (c) Li conductivities of five individual simulations with the same Cl@4c but different Li distributions with blue colors. The orange line represents the average and standard deviation of MSD obtained from 5 simulations. The relative standard deviation (RSD) is written inside the figures for each supercell.}
  \label{fgr:parameter}
\end{figure*}

In Figure 3b, MSD($t$) is displayed for a 5×5×5 supercell. A nonlinear region is noticeable for the initial period $t_{\rm NL}$, which lasts about 5 ns. 
In ref~\citenum{10.1038/s41524-018-0074-y}, the initial nonlinearity in MSD was attributed to a ballistic region where Li ions vibrate locally for a few picoseconds. 
As will be analyzed in detail in Section 3.4, the nonlinear behavior in the argyrodite systems originates from the rapid intra-cage hopping. This hopping is spatially confined within the cage and thus exhibits nonlinear behavior. After $t_{\rm NL}$, long-range diffusion, facilitated by inter-cage hopping, becomes activated, resulting in more linear curves.  
Therefore, when calculating the diffusivity at 300 K, we omit the initial 5 ns and compute $D_{\rm Li}(t)$ using linear regression of MSD($t$) over the interval $[t_{\rm NL}, t+t_{\rm NL}]$ (see Figure 3b).

By inserting $D_{\rm Li}(t)$  into eq 4, we can compute $\sigma_{\rm Li}(t)$. Figure 3c shows $\sigma_{\rm Li}(t)$ at 1-ns intervals from five runs for each supercell size, represented by solid curves. The dots and error bars indicate the average conductivity ($\overline{\sigma}_{\rm Li}(t)$) and the standard deviation among the five trajectories, respectively.
Regardless of supercell size, all $\overline{\sigma}_{\rm Li}(t)$ values converge to similar numbers. However, statistical fluctuations vary significantly depending on the supercell size.  
To investigate this in detail, we calculate the relative standard deviation (RSD), which is the standard deviation divided by $\overline{\sigma}$. 
As the supercell size increases, the RSD at $t = 20$ ns monotonically decreases from 0.238 in a 2×2×2 supercell to 0.06 in a 5×5×5 supercell. Assuming ideal random walks, one can show that the limiting RSD is $\sqrt{2/3N}$, where $N$ is the number of diffusing particles in the supercell (see Supporting Information for the derivation). Therefore, the ideal RSDs are 0.059, 0.032, 0.021, and 0.015 for supercells from 2×2×2 to 5×5×5, respectively.
However, the observed RSD at 300 K exceeds the ideal value by a factor of 4. This discrepancy arises because inter-cage hopping has not occurred sufficiently within the 25-ns simulation time such that the ideal Gaussian distribution of displacements as in the random-walk model is yet to be established. At higher temperatures such as 350 and 400 K, the RSD approaches the ideal values for the same MD simulation time (see Figure S4). Even though the ideal random-walk distribution is not achieved at 300 K,
a single MD run with a 5×5×5 supercell can still produce $\sigma_{\rm Li}$ within $\sim$10\% error margin at 95.4\% confidence (two standard deviations).  Therefore, this MD setup will be employed for subsequent computations at 300 K. We note that if one wish to use smaller supercells, it is necessary to compute the ensemble average. 

Lastly, we examine the size effect of the supercell on S/Cl distributions. To this end, we generate two additional random distributions of S/Cl in the 5×5×5 supercell with 50\% \clc . We calculate the Li conductivity  following the approach established earlier and find that the converged values differ by only 6\%. This suggests that the 5×5×5 supercell is sufficiently large  to fully accommodate the disorder of the S/Cl distribution.

\subsection{3.3. Free Energy}

\begin{figure}
  \includegraphics[width=3in]{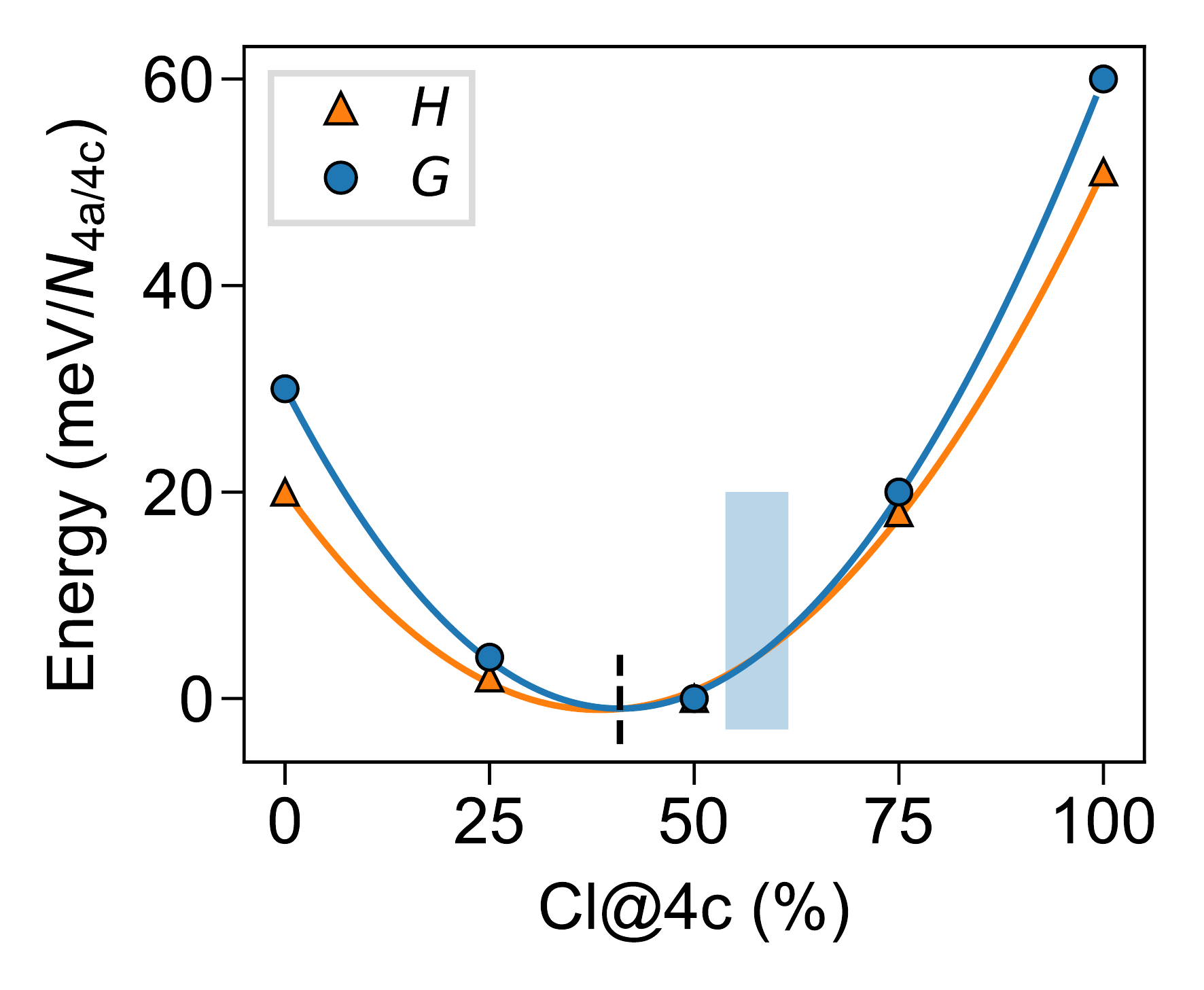}
  \caption{The enthalpy ($H$) and Gibbs free energy ($G$) for 0\%–100\% Cl@4c at 300 K, in reference to the values at 50\% Cl@4c. The curves are fitted to the energies using  third-order polynomials. The minimum point of $G$ is marked by the vertical dashed line and the semi-transparent shade represents the experimental range of Cl@4c.}
  \label{fgr:configurational}
\end{figure}

To understand the thermodynamic driving force underlying S/Cl disorder, we compute free energies with respect to Cl@4c under ambient conditions. We first evaluate the enthalpy ($H$) as the mean value during the MD simulations. The initial 5-ns period is discarded, in accordance with the discussion in the previous subsection. The lattice parameter for each MD simulation is determined using the method described in the preceding subsection. The values are 9.838, 9.816, 9.804, 9.817, and 9.848 \AA\ for 0\%, 25\%, 50\%, 75\%, and 100\% Cl@4c, respectively. Figure~\ref{fgr:configurational} displays the enthalpy relative to that of 50\% Cl@4c. The enthaply is also normalized by $N_{\rm 4a/4c}$, the total number of 4a and 4c cages that are randomly occupied by S and Cl atoms (for the unit cell, $N_{\rm 4a/4c}$ = 8).
The curve in Figure 4 is fitted to the enthalpy data using a third-order polynomial. 
The dependence of $H$ on \clc\ can be understood through the interplay between ionic size and Coulomb interaction. That is to say, the  4c cages, which are more compact than the 4a cages~\cite{10.1039/d1ta10964b} (as discussed earlier), sterically prefer smaller Cl$^{-}$ ions but electrostatically favor S$^{2-}$ ions. Balancing these two competing factors, the minimum energy is found near the middle value of \clc\ (38\% in the fitted curve).

Next, we evaluate the Gibbs free energy ($G$) that defines the equilibrium state. In the present study, we do not account for the order-disorder transition, which would require a statistical model to accurately calculate the enthalpy for specific S/Cl distributions. While this is feasible using NNP, we defer this aspect to future studies. Instead, we assume that S/Cl is fully disordered for a certain value of \clc\ and aim to determine the equilibrium site occupancy. We further assume that the vibrational entropy is similar across all cases. The configurational entropy from S/Cl disorder can be calculated using the 
binary cross-entropy formula for each 4a and 4c site. If the site occupancy of S or Cl at either 4a or 4c sites is $p$, the total configurational entropy induced by disorder ($S_{\rm conf}$) is given by the following expression:

\begin{equation}
S_{\rm conf}=-k_{\rm B}N_{\rm 4a/4c} 
\left[ p \ln{p}  + (1-p) \ln{(1-p)} \right]
\label{eqn:entropy}
\end{equation}

The Gibbs free energy ($G$) is obtained as follows:

\begin{equation}
G = H - T S_{\rm conf}
\end{equation}

Figure~\ref{fgr:configurational} shows $G$ relative to that for 50\% \clc\ at 300 K and a third-order polynomial fitted to $G$. Since the entropy is the maximized at 50\%, the minimum point of $G$ shifts closer to the middle value compared to $H$, as indicated by the vertical dashed line (41\%). The theoretical site occupancy at equilibrium aligns reasonably well with the experimental range of 53.8\%--61.5\%\ (see the shaded region), corresponding to a slightly higher $G$ ($\sim$5 meV/$N_{\rm 4a/4c}$). In ref~\citenum{10.1021/acs.chemmater.0c04650}, a small disorder level of 10\% was achieved by annealing at low temperatures, which suppresses the entropic driving force towards higher levels of disorder.

\subsection{3.4. Ionic Conductivity and Activation Energy}

\begin{figure*}
  \includegraphics[width=6in]{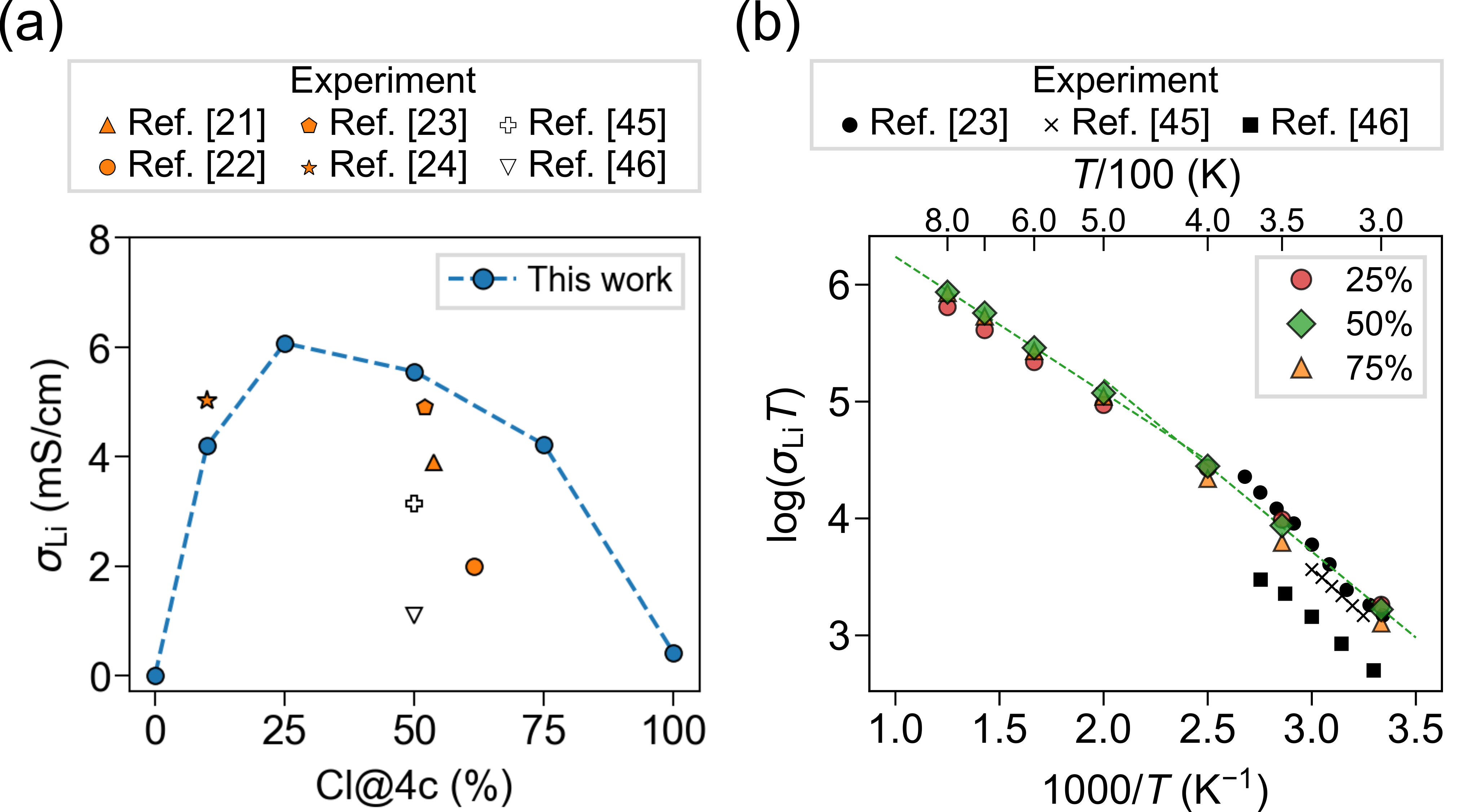}
  \caption{(a) Li conductivity at 300 K as a function of the Cl occupancy on the 4c site. The orange, open, and blue symbols represent the values from experiments with reported Cl@4c, without Cl@4c, and this work, respectively. (b) Arrhenius plots of Li conductivity for 25\%, 50\%, and 75\% Cl@4c. Data points are obtained at 300, 350, 400, 500, 600, 700, and 800 K. The red, green, orange, and black markers represent 25\%, 50\%, 75\% Cl@4c, and experimental values, respectively. The green dashed lines are fitted at high (500-800 K) or low temperatures (300-400 K) for 50\% \clc.}
  \label{fgr:ionic}
\end{figure*}

With the computational parameters established in Section 3.2, we calculate the Li-ion conductivity of \ch{Li6PS5Cl} at  0\%, 10\%, 25\%, 50\%, 75\% and 100\% Cl@4c at 300 K. The results are shown in Figure~\ref{fgr:ionic}a along with experimental data. The computed conductivities are 0.0, 4.2, 6.1, 5.5, 4.2, and 0.4 mS/cm for 0\%, 10\%, 25\%, 50\%, 75\%, and 100\% Cl@4c, respectively. As expected, the conductivity of the disordered structures is significantly higher than that of the ordered structures. It is intriguing that $\sigma_{\rm Li}$ peaks at  25\%,  rather than at 50\% where the disorder is maximized. This is at variance with ref 41 which suggested the maximum $\sigma_{\rm Li}$ at 75\%. A detailed analysis will be provided in the below. 

Experimental values obtained at structures with 52\%–62\% Cl@4c range from 2 to 5 mS/cm,\cite{10.1021/acs.chemmater.0c02418, 10.1021/jacs.7b06327, 10.1021/acsami.8b07476} which agree reasonably well with our results. (The open symbols in Figure 5a are taken from  experiments where \clc\ was not specified.)
We note that in ref~\citenum{10.1021/acsami.8b07476},  \ch{Li6PS5Cl} was synthesized by direct annealing of mixed precursors, avoiding intensive ball milling. This approach likely increased the grain size and reduced defect concentrations compared to other experiments. Notably, the conductivity in this sample is closest to our computed value, which was obtained for a single crystal without any defects.  

Figure~\ref{fgr:ionic}b shows the Arrhenius plot for 25\%, 50\%, and 75\% Cl@4c from 300 to 800 K (see eq 6). The conductivities at 350--800 K are obtained from a single MD calculation with 4×4×4 supercell for a total of 25 ns. This computational setup was determined by following the same procedure for 300 K in Section 3.2. In detail, we carry out MD simulations with 50\% Cl@4c at 350 and 400 K (see Figure S4a). The linearity of MSD($t$) becomes evident after 2 and 1 ns for 350 and 400 K respectively, indicating the acceleration of long-range diffusion compared to the room temperature. Figure S4b shows that a single calculation with supercells $\geq$ 3×3×3 produce a converged $\sigma_{\rm Li}$ within the error range of 10\%, and we select 4×4×4 supercell which is computationally still affordable. We apply the same setting throughout temperatures higher than 300 K. 

In Figure 5b, we note that the slope changes slightly around 500 K (see dashed lines for 50\% \clc). 
At high temperatures, $E_{\rm a}$ is 0.222, 0.231, and 0.236 eV, and it increases to 0.281, 0.292, and 0.294 eV at low temperatures for 25\%, 50\%, and 75\% Cl@4c, respectively. The experimental $E_{\rm a}$ measured over 300–350 K is 0.27–0.34 eV,\cite{10.1021/acsami.8b07476, 10.1021/acsami.8b15121, 10.1149/2.0301903jes} which agree well with this work. 
As a computational point, we remark that due to the changes in $E_{\rm a}$, extrapolating from high temperatures may lead to inaccuracies. For instance, the extrapolated value for 50\% \clc\ is 11.1 mS/cm at 300 K, which is twice the actual value (5.5 mS/cm). This suggests that direct simulations at ambient temperatures are necessary to obtain quantitatively accurate results.

To understand the disorder-dependent conductivity, in particular asymmetric behavior, we analyze the diffusional motion of Li by calculating the self part of the van Hove correlation function ($G_{\rm{s}}$):\cite{10.1103/PhysRev.95.249}

\begin{equation}
G_{\rm{s}}(r,t) = \frac{1}{4\pi r^2N_{\rm{Li}}}\left\langle\sum_{i=1}^{N_{\rm{Li}}}\delta(r-\left|\mathbf{r}_{i}(t_{0})-\mathbf{r}_{i}(t+t_{0})\right|)\right\rangle_{t_{0}}
\end{equation}
where  $\delta$ is the Dirac delta function and $\left\langle \cdot \right\rangle_{t_{0}}$ represents averaging over the initial time $t_{0}$. $G_{\rm{s}}(r,t)$ represents the probability of finding a Li-ion at the distance $r$ and time $t$ from its initial position, which informs about Li-hopping behaviors. 
The van Hove correlation function has been applied to various solid-state electrolytes to understand the jumping dynamics of ions and reveal the diffusion mechanism.\cite{10.1038/ncomms15893, 10.1021/acs.chemmater.5b03656, 10.1039/d1ta10964b, 10.1021/acsami.6b00833, 10.1021/acs.chemmater.6b02648}
Figure 6 depicts $G_{\rm{s}}(r,t)$ for each \clc . As indicated at the side of Figure 6c, $0 < r < 1 \text{ \AA}$ corresponds to Li atoms fixed at their initial positions with small vibrations. Given that the diameters of the 4a or 4c cages are approximately 5 \AA, Li atoms at $1 < r < 5 \text{ \AA}$ have mostly undergone only intra-cage jumps, while those with $ r > 5 \text{ \AA}$ have escaped the original cage via inter-cage jumps.

\begin{figure*}
  \includegraphics[width=6in]{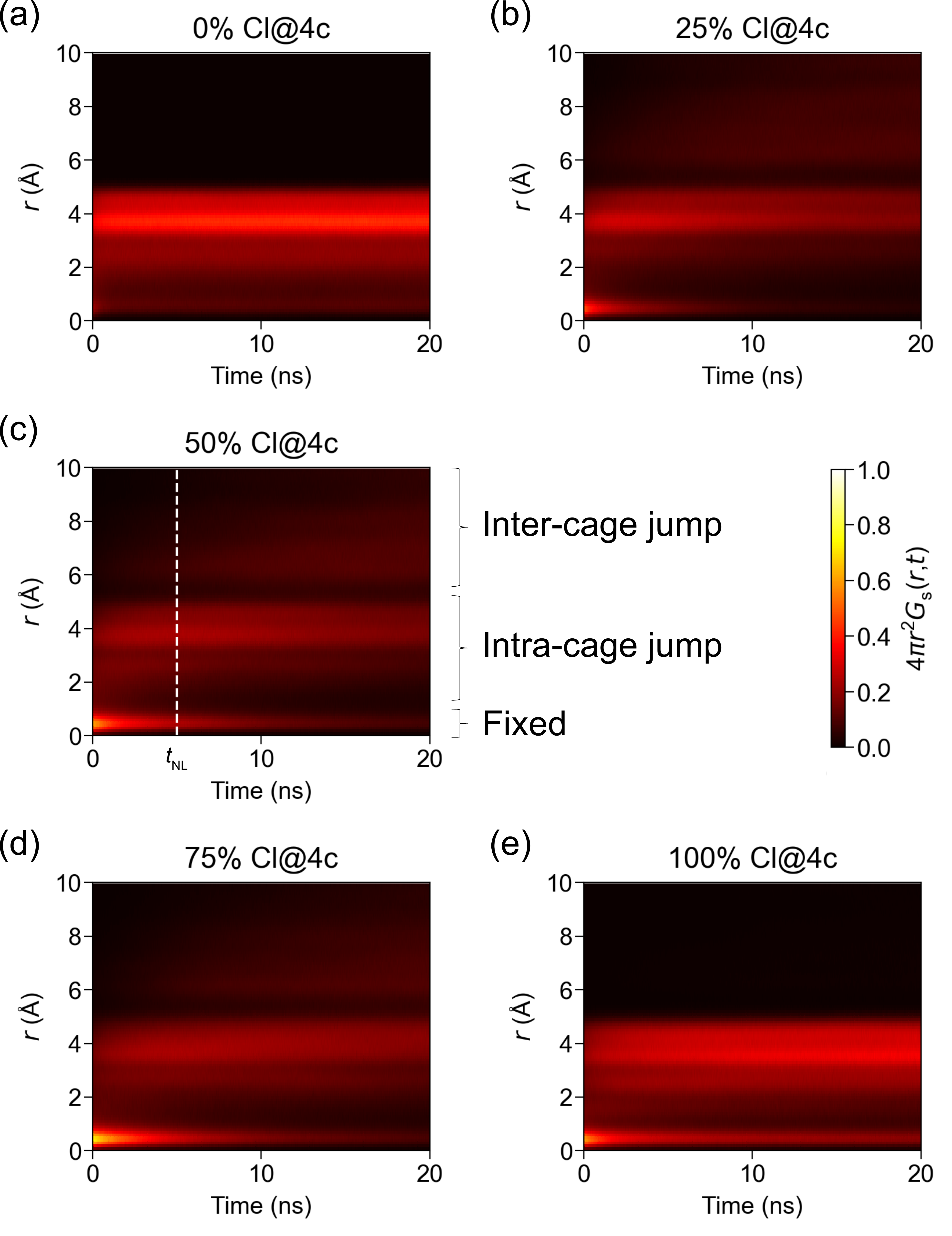}
  \caption{Self part of the van Hove correlation function of Li with (a) 0\%, (b) 25\%, (c) 50\%, (d) 75\%, and (e) 100\% Cl@4c obtained from MD simulations at 300 K. The black and white indicate the lowest and highest data point density, respectively.}
  \label{fgr:vanHove}
\end{figure*}

In Figures 6a and 6e, S atoms fully occupy the 4c and 4a cages, respectively, and Li atoms effectively form the corresponding cages as they are attracted to S atoms (see Figure 2b). In both figures, inter-cage jumps are negligible, which is consistent with the small conductivity for the ordered structures. In Figure 6a, intra-cage jumps are dominant and Li atoms primarily move to other sites in the same cage within 1 ns. This contrasts with Figure 6e in which a significant portion of Li atoms remain fixed to their original positions. This can be understood as follows: In the case of the 4c cage, the low-barrier shuttling motion along the 48h-24g-48h dimer (see Figure 1a) facilitates active rotation of Li atoms within the 4c cages. However, in the 4a cage, such a low-barrier path is absent, and considerable atoms remain fixed for a prolonged period.

For intermediate \clc\ as shown in Figures 6b-d, Li diffusion is percolated throughout the material as S atoms occupy both 4a and 4c sites, yielding finite probabilities for inter-cage jumps. 
In Figure 6c, the inter-cage jump starts near 5 ns or $t_{\rm NL}$, which is consistent with the time for showing non-linearity in MSD (see Figure 3b).
The degree of disorder is at its maximum at 50\% \clc, but the conductivity for 25\% \clc\ is higher (see Figure 5a) because intra-cage jumps are more active than at 50\% \clc: The enhanced rotational motion within the cage should allow more Li atoms to access the percolated inter-cage pathway. This explains the asymmetric behavior of Li conductivity  in Figure 5a. 
That is to say, two primary factors contribute to enhancing Li conductivity in \ch{Li6PS5Cl}. The first one is the disorder effect, which percolates the diffusion pathway between 4a and 4c cages and is maximized at 50\% Cl@4c. The second one is the fast rotational motion that enables more Li atoms to be involved in long-range diffusion, and this is enhanced as Cl@4c approaches 0\%. Therefore, the optimal Cl@4c value, considering both effects, lies between them; in our case, the conductivity peaks at 25\% Cl@4c.
From a thermodynamic perspective, 25\% \clc\ has a value of 3.6 meV/$N_{\rm 4a/4c}$ (see Figure 4), making it thermodynamically accessible and thus synthesizable.

On the other hand, the nonlinear behavior in $\log (\sigma_{\rm Li}T)$ vs. $1/T$ plot (see Figure 5b) could be attributed to the enhanced rotation frequency of Li ions in 4a cages at elevated temperatures. This is supported by much higher values of $G_{\rm{s}}(r,t)$ at $1 < r < 5 \text{ \AA}$ in a 100\% Cl@4c model at 500 K (Figure S5). Consequently, other effective routes for Li-ion hopping, which rarely appears at low temperatures, can be open at high temperatures, reducing $E_{\rm a}$ as confirmed in the simulation. In addition, the facile Li-ion rotation in both 4a and 4c cages leads to uniform diffusion pathways at high temperatures (Figure S6). Due to this fact, $\sigma_{\rm Li}$ is higher at 50\% Cl@4c than 25\% or 75\% at temperatures of 500--800 K (see Figure 5b).

To note, a monotonic increase in $\sigma_{\rm Li}$ was observed in experiment for \ch{Li6PS5Br} as disorder increases, \cite{10.1002/aenm.202003369} unlike the asymmetric behavior in \ch{Li6PS5Cl} predicted in the present study. For \ch{Br-}, the increase in lattice softness may promote intra-cage hopping, while larger inter-cage distances due to increased lattice parameters make the inter-cage jump as a rate-determining step. Thus, an increase in disorder can lead to a more homogeneous distribution of Li-ions, thereby facilitating inter-cage jumps \cite{10.1021/acsami.8b07476}.

\section{CONCLUSION}

In conclusion, we developed an accurate NNP and used it to perform MD simulations on disordered \ch{Li6PS5Cl} over a temperature range of 300–800 K, yielding insights into Li-ion diffusion.  Our analysis rigorously tested statistical parameters to achieve a conductivity at 300 K that converged within an error range of approximately 10\%. 
As a result, we employed 25-ns simulations using a 5×5×5 supercell containing 6,500 atoms. The calculated Li-ion conductivity, activation energies, and equilibrium \clc\ are in good agreement with experimental data. 
Notably, the peak in Li conductivity is observed  when Cl ions occupy 25\% of 4c sites, rather than at 50\%, where disorder is maximized. This can be attributed to the interplay between inter-cage and intra-cage jumps.
Considering that \clc\ of 10\% was achieved experimentally through annealing at low temperatures~\cite{10.1021/acs.chemmater.0c04650} and high-temperature annealing followed by rapid cooling enabled tunability of site disorder in \ch{Li6PS5Br}~\cite{10.1002/aenm.202003369},  we suggest that \ch{Li6PS5Cl} with $\sim$25\% \clc\ could also be synthesized, potentially boosting Li-ion conductivity to near 10 mS/cm.  
To emphasize, the detailed analysis of the diffusion mechanism in \ch{Li6PS5Cl} presented in this study was made possible by large-scale MD simulations at room temperature, enabled by our accurate NNP. By shedding light on key factors affecting Li-ion diffusion in \ch{Li6PS5Cl}, we believe this work lays the foundation for optimizing ionic conductivity in the argyrodite family.

\section{ASSOCIATED CONTENT}

\begin{suppinfo}
The following files are available free of charge.
\begin{itemize}
  \item Supporting Information.pdf: Derivation of ideal relative standard deviation for the diffusion coefficient, parity plot of energy and forces, validations of NNP, parameter tests for high temperatures, $G_{\rm{s}}(r,t)$ for 100\% Cl@4c at 500 K, Li trajectories at high temperatures, and details for NNP training.
  \item Movies.zip: Time-averaged movies of 0\%, 25\%, 50\%, 75\%, and 100\% Cl@4c 25-ns NNP-MD trajectories with 5$\times$5$\times$5 supercells and averaging window of 200 ps.
\end{itemize}
\end{suppinfo}

\section{AUTHOR INFORMATION}

\subsection{Corresponding Author}

\begin{itemize}
    \item Seungwu Han – Department of Materials Science and Engineering, Seoul National University, Seoul 08826, Korea; Korea Institute for Advanced Study; orcid.org/0000-0003-3958-0922; Email: hansw@snu.ac.kr
\end{itemize}

\subsection{Authors}

\begin{itemize}
    \item Jiho Lee – Department of Materials Science and Engineering, Seoul National University, Seoul 08826, Korea; orcid.org/0009-0008-7266-091X
    \item Suyeon Ju – Department of Materials Science and Engineering, Seoul National University, Seoul 08826, Korea; orcid.org/0000-0001-7033-8769
    \item Seungwoo Hwang – Department of Materials Science and Engineering, Seoul National University, Seoul 08826, Korea; orcid.org/0000-0002-1523-8340
    \item Jinmu You – Department of Materials Science and Engineering, Seoul National University, Seoul 08826, Korea; orcid.org/0009-0003-7152-7774
    \item Jisu Jung – Department of Materials Science and Engineering, Seoul National University, Seoul 08826, Korea; orcid.org/0000-0003-2814-1289
    \item Youngho Kang – Department of Materials Science and Engineering, Incheon National University, Incheon 22012, Korea; orcid.org/0000-0003-4532-0027
\end{itemize}

\subsection{Author Contributions}

Jiho Lee and Suyeon Ju contributed equally to this work.

\subsection{Notes}

The authors declare no competing financial interest.

\begin{acknowledgement}
This work was supported by the National Research Foundation of Korea (NRF) grant funded by the Korea government (MSIT) (00247245). The computations were carried out at the Korea Institute of Science and Technology Information (KISTI) National Supercomputing Center (KSC-2023-CRE-0337).
\end{acknowledgement}

\bibliography{references}

\end{document}